\documentclass{nime-alternate} %

\usepackage{anonymize} 		   %

\usepackage[utf8]{inputenc}
\usepackage{booktabs}
\usepackage{multirow}
\usepackage{threeparttable}

\usepackage{lineno}

\begin{document}%

\conferenceinfo{NIME'24,}{4--6 September, Utrecht, The Netherlands.}

\title{Real-time Timbre Remapping with Differentiable DSP}

\label{key}%

\numberofauthors{4} %
\author{
\alignauthor
\anonymize{Jordie Shier}\\
       \affaddr{\anonymize{Centre for Digital Music}}\\
       \affaddr{\anonymize{Queen Mary University of London, UK}}\\
       \email{\anonymize{j.m.shier@qmul.ac.uk}}
\alignauthor
\anonymize{Charalampos Saitis}\\
       \affaddr{\anonymize{Centre for Digital Music}}\\
       \affaddr{\anonymize{Queen Mary University of London, UK}}\\
       \email{\anonymize{c.saitis@qmul.ac.uk}}
\alignauthor
\anonymize{Andrew Robertson}\\
       \affaddr{\anonymize{Ableton AG}}\\
       \affaddr{\anonymize{Berlin, Germany}}\\
       \email{\anonymize{andrew.robertson@ableton.com}}
\and  %
\alignauthor
\anonymize{Andrew McPherson}\\
       \affaddr{\anonymize{Dyson School of Design Engineering}}\\
       \affaddr{\anonymize{Imperial College London, UK}}\\
       \email{\anonymize{andrew.mcpherson@imperial.ac.uk}}
}

\maketitle

\begin{abstract}\sloppy
Timbre is a primary mode of expression in diverse musical contexts.
However, prevalent audio-driven synthesis methods predominantly rely on pitch and loudness envelopes, effectively flattening timbral expression from the input.
Our approach draws on the concept of timbre analogies and investigates how timbral expression from an input signal can be mapped onto controls for a synthesizer.
Leveraging differentiable digital signal processing, our method facilitates direct optimization of synthesizer parameters through a novel feature difference loss. This loss function, designed to learn relative timbral differences between musical events, prioritizes the subtleties of graded timbre modulations within phrases, allowing for meaningful translations in a timbre space.
Using snare drum performances as a case study, where timbral expression is central, we demonstrate real-time timbre remapping from acoustic snare drums to a differentiable synthesizer modeled after the Roland TR-808.

\end{abstract} 



\section{Introduction}
Timbre is a musical concept that has distinctly resisted precise definition in psycoacoustics and music psychology research \cite{krumhansl_why_1989}.
It has been referred to as the ``psychoacoustician's multidimensional waste-basket category for everything that cannot be labeled pitch or loudness” \cite[p.34]{mcadams_hearing_1979}.
Yet, within that multidimensional waste-basket resides a rich landscape of musical expression.
Timbre is central to many musical traditions including Western classical music \cite{mcadams_timbre_2019}, electronic music \cite{reynolds_energy_2013}, and diverse percussion traditions including tabla \cite{chordia_segmentation_2005} and drum kit performance \cite{danielsen_effects_2015}.
The sound synthesizer has played a particularly important role in expanding the timbral palette of musicians and entirely new musical cultures have formed around their use \cite{bates_interface_2021}.
Timbre perception and synthesizers share a rich history, evidenced by the significant body of literature following Wessel's introduction of the timbre space as a control interface in 1979 \cite{wessel_timbre_1979}.
Furthermore, deep learning has enabled novel timbre-focused synthesis tasks including timbral control \cite{roche_make_2021} and musical timbre transfer \cite{huang_timbretron_2019, engel_ddsp_2020}.

Timbre transfer has received considerable attention in recent years and refers to the task of altering ``timbral information in a meaningful way while preserving the hidden content of performance control" \cite[p.4]{dai_music_2018}.
One popular formulation of this task was introduced by Engel et al. \cite{engel_ddsp_2020} and involves learning a mapping from pitch and loudness control signals to the timbre of a target instrument, expressed as time-varying harmonic amplitudes.
In other words, pitch and loudness are explicit control signals and timbre is implicitly learned, conditional on time-varying pitch and loudness.
The side-effect of this technique is that timbral expression is effectively ignored in the input signal.
Whilst this may suit musical contexts where pitch and loudness are the primary parameters of expression, what about the myriad contexts where this is not the case?
In this paper we propose a novel, data-driven formulation for timbral control of synthesizers using differentiable digital signal processing (DDSP) \cite{engel_ddsp_2020}, aiming to address musical contexts where timbre is a primary vehicle for musical expression.

We draw on the idea of timbre remapping, introduced by Stowell and Plumbley \cite{stowell_timbre_2010}, who defined the task as mapping trajectories between two distinct timbre spaces.
This concept bears similarity to the idea of timbre analogies \cite{wessel_timbre_1979, mcadams_perception_1992}, that is, transpositions of sequences within a timbre space, and provides a conceptual starting point for our design.

Instead of learning to match absolute audio feature values, as is typical in AI-based audio synthesis, we propose to learn to match relative differences in audio features.
This design decision is motivated by the role of timbre in the structuring of musical phrases \cite{mcadams_timbre_2019} and the importance of subtle, graded timbral differences \cite{wessel_control_1987}.
It is the relationship between the timbres of neighboring musical events that is important in the creation of a musical phrase, not the absolute values of each individual event taken in isolation.
This is exemplified by drummers intentionally varying the intensity and timbre of certain hits \textit{within} a groove to provide juxtaposition and indicate rhythmic intention \cite{danielsen_effects_2015}.

To this end, we present a \textit{feature difference loss} function that considers pairs of sounds and differences between their features as opposed to absolute values in isolated notes.
Paired with a differentiable synthesizer and a gradient-based optimizer, this enables us to learn to adjust synthesizer parameters to create timbral differences analogous to those observed between successive events in a musical passage -- remapping the timbre from a musical passage onto a synthesizer.

As a case study, we consider the musical context of a snare drum performance \cite{danielsen_effects_2015} and demonstrate how this method can be used for timbre remapping from acoustic snare drums to a differentiable synthesizer modeled after the Roland TR-808 snare drum.
An open source audio plug-in implementing the real-time system and  training scripts are presented alongside this paper to allow musicians to experiment in their own musical contexts.
Reflections on a session with a professional drummer are provided and point to both the effectiveness of this approach as well as areas for future improvement.
Recordings from this session and software are available on a supplementary website\footnote{\url{https://jordieshier.com/projects/nime2024/}}.

\section{Background}
The work presented in this paper fits into the broader landscape of work focused on the development of novel controllers and mappings for synthesizers \cite{miranda_new_2006}.
It also contributes to the growing body of literature in NIME on machine learning for musical expression \cite{jourdan_machine_2023}.

\subsection{Timbre Spaces and Timbre Analogies}
\label{sec:background-timbre}
The perceptual foundation of this work is the timbre space, a multi-dimensional representation of sounds, derived from listening studies on perceived similarity \cite{grey_multidimensional_1977} and the lineage of work that followed, seeking acoustic correlates with timbre perception \cite{grey_perceptual_1978} (see \cite{mcadams_perceptual_2019} for an overview).
The MPEG-7 standard \cite{isoiec_isoiec_2002} defined a set of audio features to quantify timbre, although a multitude of other features have been proposed \cite{peeters_large_2004} and are commonly used.
The concept of timbre analogies was first explored by Ehresman and Wessel in 1978 \cite{ehresman_perception_1978} and described transpositions of sequences within a perceptual timbre space.
Wessel subsequently discussed timbre analogies in the context of musical control with timbre spaces \cite{wessel_timbre_1979} and McAdams and Cunible verified the perceptual viability of timbre analogies \cite{mcadams_perception_1992}.
Perceptual studies performed by these researchers utilized pairs of stimuli and asked whether participants were sensitive to relative differences in timbre.
More concretely, given a pair of sounds $\mathbf{x}_a$ and $\mathbf{x}_b$, participants were asked to select a sound $\mathbf{x}_d$ (from a set of choices) that differed from $\mathbf{x}_c$ by the same amount as $\mathbf{x}_b$ differed from $\mathbf{x}_a$.
Results showed that $\mathbf{x}_d$ could be predicted by a parallelogram model of similarity within a perceptual timbre space.
From a geometric perspective, this indicates that timbral sequences could be represented as vectors in a multidimensional timbre space and translated within that space while preserving the perception of relative difference.
While the perception of these relative differences was not as strong as pitch and depended on the nature of the relationship \cite{mcadams_perception_1992}, it suggests that with the correct perceptual scalings, translations of timbre are viable musical operations -- an idea we build upon in this work. 

\subsection{Perceptual Control of Synthesizers}

Fasciani defines a synthesis method as being perceptually related when it explicitly manipulates timbral attributes of the generated sound \cite{fasciani_interactive_2020}.
Timbre remapping can be considered an example of a perceptual control method as it seeks to explicitly manipulate timbre by mapping attributes from an input source to the generated sound.
Stowell and Plumbley \cite{stowell_timbre_2010} introduced this concept and identified the difficultly of the task when the distribution of timbres differ between the control and target contexts.
As a practical example, they presented the task of controlling a concatenative synthesizer using an audio signal, building on the work of Schwarz \cite{schwarz_data-driven_2004}.
A key contribution by Stowell and Plumbley \cite{stowell_timbre_2010, stowell_learning_2011}, \cite[Chapter 5]{stowell_making_2010} is the recognition of the context-dependent nature of timbre and multidimensional interactions between various features.
They proposed an unsupervised regression tree method to learn associations between the distinct timbre spaces of the input control and synthesizer.

Another line of work considers the timbre space directly as musical control structure.
Building on Wessel's 1979 paper \cite{wessel_timbre_1979}, several researchers have explored computational methods for navigating timbre spaces with respect to synthesis parameters \cite{hoffman_feature-based_2006-1, gregorio_augmenting_2019, fasciani_interactive_2020, sramek_soundtraveller_2023, vaillant_interpolation_2023}.
Timbral exploration methods are often motivated to cover the space of all possible sounds of a synthesizer to support searching; however, this can make subtle timbral variations more challenging to achieve \cite{gregorio_augmenting_2019}.
It is these subtle timbral variations that we turn our attention to in this work -- ``graded timbral differences" that contribute to the perception of a continuous musical phrase \cite{wessel_control_1987}.

\subsection{Differentiable Digital Signal Processing}
Differentiable digital signal processing (DDSP) was introduced by Engel et al. alongside an implementation of a harmonic plus noise synthesizer for modeling monophonic and harmonic instruments \cite{engel_ddsp_2020}.
DDSP enables the integration of DSP algorithms directly into neural network training regimes, allowing for loss functions to be computed directly on generated audio as opposed to parameter values, better representing the complex relationship between the auditory and parameter space \cite{esling_flow_2020}.
Following the initial DDSP paper, a large body of work on audio synthesis has followed exploring numerous synthesis methods including waveshaping \cite{hayes_neural_2021}, FM \cite{caspe_ddx7_2022, yang_white_2023}, subtractive \cite{masuda_improving_2023}, and filtered noise synthesis \cite{barahona-rios_noisebandnet_2024}.
Of particular relevance is the DDSP timbre transfer task \cite{engel_ddsp_2020, carney_tone_2021}, which follows naturally from the choice of pitch and loudness as control signals.
The timbre of an instrument learned during training, represented by time-varying harmonic amplitudes, can be mapped onto pitch and loudness contours of a different instrument during inference.
In contrast to the DDSP timbre transfer formulation, we propose a method that considers how timbre can be explicitly used as a control signal.
For a full review of DDSP for audio synthesis see Hayes et al. \cite{hayes_review_2024}.

\section{Timbre Remapping Approach}

Here we outline the design of a timbre remapping approach, building on the concept of timbre analogies within the framework of DDSP.
The goal is to transform the timbre of a target synthesized sounds by mapping specific dimensions of a performance from an input control signal, which we assume includes variations in timbre that can be measured using acoustic features.
We also assume that the input and synthesized sounds occupy unique regions within a multidimensional timbre space.
Based on research on timbre analogies introduced in section \ref{sec:background-timbre}, we propose timbre remapping via translation within a computational timbre space.
The basic idea is to measure how timbre changes across successive musical events in an input audio control signal and translate those changes into synthesizer parameter modulations.
In the next section we outline a method for performing this using DDSP.

\begin{figure}[t]
	\centering
	\includegraphics[width=0.47\textwidth]{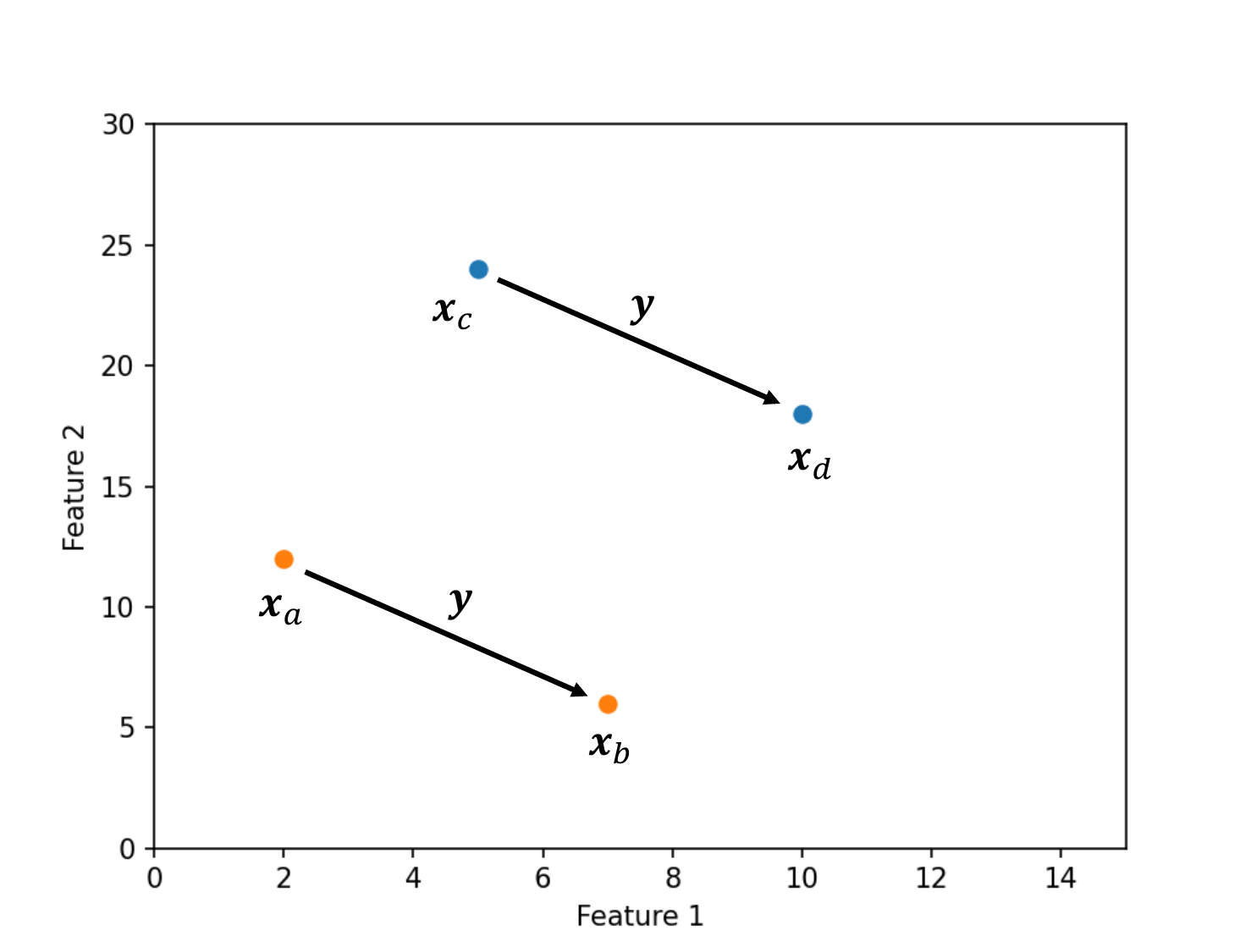}
	\caption{Two-dimensional representation of a timbre analogy. The sound pair $(\mathbf{x}_a, \mathbf{x}_b)$ form a timbre sequence and differ by $\mathbf{y}$. Given a new sound $\textbf{x}_c$ (e.g., a synthesizer sound), an analogous timbre sequence can be created by applying the difference described by $\textbf{y}$ to form the new pair $(\mathbf{x}_c, \mathbf{x}_d)$.}
	\label{fig:timbre_analogy}
\end{figure}

\subsection{Learning Feature Differences}

We start with a timbral sequence comprising two sounds $\mathbf{x}_a$ and $\mathbf{x}_b$ from an input control source.
A multidimensional timbre space is defined by an arbitrary audio feature extraction algorithm $f(\cdot)$ that returns a multidimensional vector of features.
The timbral sequence can be described by the vector resulting from taking the difference between audio feature vector of $\mathbf{x}_a$ and $\mathbf{x}_b$:

\begin{equation}\label{eqn:ref}
	\mathbf{y} = f(\mathbf{x}_{b}) - f(\mathbf{x}_{a})
\end{equation}

Now, given a synthesizer $g(\cdot)$ and a preset $\mathbf{\theta}_{pre} \in \mathbb{R}^P$, where $P$ is the number of synthesizer parameters, we can synthesize an audio signal $\mathbf{x}_c = g(\mathbf{\theta}_{pre})$.
We can also generate a modulated version of that preset by applying a parameter modulation $\mathbf{\theta}_{mod} \in \mathbb{R}^P$ which results in a new synthesized signal $\mathbf{x}_d = g(\mathbf{\theta}_{pre} + \mathbf{\theta}_{mod})$.
This forms a new timbre sequence:

\begin{align}\label{eqn:yhat}
	\mathbf{\hat{y}} &= f(\mathbf{x}_{d}) - f(\mathbf{x}_{c}) \\
	&= f\left(g(\mathbf{\theta}_{pre} + \mathbf{\theta}_{mod})\right) - f\left(g(\mathbf{\theta}_{pre})\right)
\end{align}

Our goal is to learn $\mathbf{\theta}_{mod}$ such that $\hat{y} = y$. Figure \ref{fig:timbre_analogy} shows a visual overview of this process.
In this formulation, $\mathbf{\theta}_{pre}$ and $\mathbf{x}_{a}$ are fixed and can be selected based on the musical application.
To situate this formulation within a gradient descent-based machine learning paradigm, we introduce a loss function to optimize $\mathbf{\theta}_{mod}$ to match feature differences.

\subsubsection{Feature Difference Loss}
\label{sec:feat-diff-loss}

DDSP synthesizer training involves optimizing the parameters of synthesizer (and optionally a neural network that predicts parameters) to minimize a loss function.
Typically, loss is computed between predicted audio and ground truth audio using an auditory loss function such as the multi-scale spectral loss \cite{wang_neural_2019}.
This formulation minimizes the absolute error between spectrograms with the objective of replicating ground truth audio.
However, we are interested in optimizing synthesizer parameters to match a difference in audio features instead of matching the absolute values of features.
To this end, we define a \textit{feature difference loss}:

\begin{equation}
    \mathcal{L}(\mathbf{\hat{y}}, \mathbf{y}) = \left\lVert \mathbf{\hat{y}} - \mathbf{y} \right\rVert_{1}
\end{equation}

\noindent where $\mathbf{\hat{y}}$ is the feature difference vector from equation \ref{eqn:yhat}, $\mathbf{y}$ is the reference difference vector from equation \ref{eqn:ref}, and $\left\lVert \cdot \right\lVert_{1}$ is the $L_1$ norm.

\subsubsection{Optimization Target}
If both $g(\cdot)$ and $f(\cdot)$ are differentiable functions then $\mathbf{\theta}_{mod}$ can be directly optimized using gradient descent. Putting this all together, we arrive at a final optimization target:

\begin{align}
	\hat{\mathbf{\theta}}_{mod} = \text{argmin}_{\theta_{mod}\in\mathbb{R}^{P}}\mathcal{L}(\mathbf{\hat{y}}, \mathbf{y})
\end{align}

This formulation makes no assumptions regarding the nature of $g(\cdot)$ and $f(\cdot)$ and only requires differentiability, which is relatively straightforward to achieve given the maturity of modern auto-differentiation software\footnote{For example, PyTorch \url{https://pytorch.org/} and TensorFlow \url{https://www.tensorflow.org/}}. In the next section we provide a concrete example using this formulation.

\section{Case Study: Snare Drums}
As a case study we investigate the application of timbre analogies and DDSP for timbre remapping within the musical context of snare drum performances.
The design of this study is motivated by prior work by Danielsen et al. \cite{danielsen_effects_2015}, which investigated the role of dynamic and timbral variation on snare drums within drum kit performances.
They found that drummers systematically varied intensity and timbre of snare drum hits within grooves to signal rhythmic and timing intentions.
In this case study, we explore how variations in a snare drum performance can be mapped onto parameters of a snare drum synthesizer.
Our goal is to perform an initial evaluation of the proposed approach for modeling timbre variation and to demonstrate a practical example supporting real-time music interaction.

\subsection{Differentiable Drum Synthesizer}

\begin{figure}[t]
	\centering
        \includegraphics[width=0.46\textwidth]{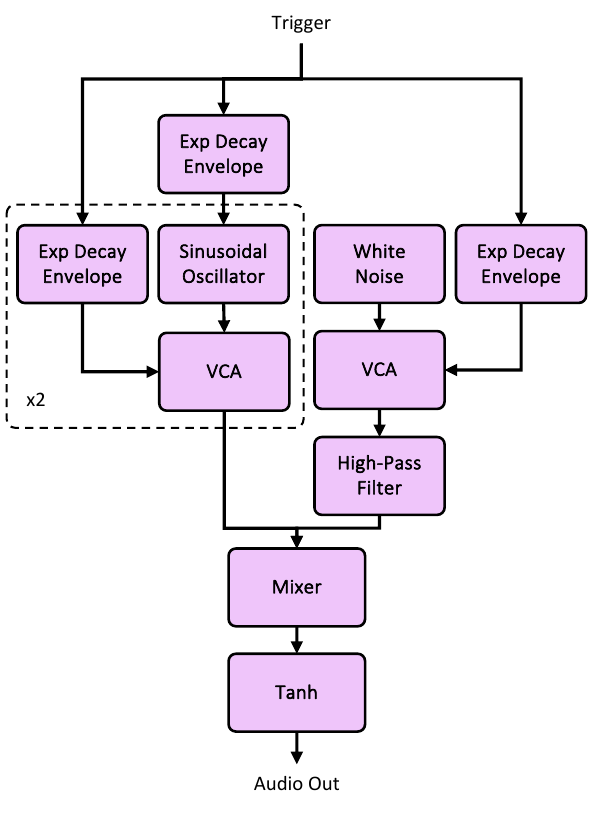}
	\caption{Drum synthesizer block diagram, modeled after the Roland TR-808 snare drum.}
	\label{fig:synth_diagram}
\end{figure}

The design of our differentiable drum synthesizer $g(\cdot)$ is inspired by the popular Roland TR-808 snare drum.
Although it is relatively simple in design, the Roland TR-808 has found widespread use within popular music~\cite{hasnain_how_2017}.
We implemented a modified version of the TR-808 snare model based on schematics provided by Gordon Reid~\cite{reid_practical_2002-1}.
A block diagram of the synthesis model is shown in figure \ref{fig:synth_diagram}.

This synthesizer consists of two parallel paths, the first comprises a pair of sinusoidal oscillators to generate the main resonant frequencies and the second is a noise generator responsible for the sound of the ``snares".
Each sinusoidal oscillator has a frequency parameter and a parameter to control amount of frequency modulation applied by a control envelope.
All sound sources have an independent amplitude control with gain and an envelope.
Frequency and amplitude envelopes are exponentially decaying envelopes with control over the decay time.
The noise source is filtered with a high-pass biquad filter and uses the differentiable implementation introduced by Yu and Fazekas \cite{yu_singing_2023}.
We added a hyperbolic tangent waveshaper to the output as we found the ability to add harmonics beneficial for shaping transients in addition to being aesthetically pleasing.
There are fourteen synthesis parameters in total.

\subsection{Audio Features}

Next, we define a feature extraction algorithm $f(\cdot)$ for measuring dynamic and timbre variations.
The exact definition of $f(\cdot)$ is not fixed and can be designed based on musical context.
Here, we select features based on Danielsen et al. \cite{danielsen_effects_2015}, who used sound pressure level, temporal centroid, and spectral centroid.
Their selection was motivated by the MPEG-7 standard \cite{isoiec_isoiec_2002} and previous research on percussive timbre analysis \cite{lakatos_common_2000, pampalk_computational_2008}.
We use this set of features and also include spectral flatness \cite{dubnov_generalization_2004} based on it's inclusion in SP-Tools\footnote{\url{https://github.com/rconstanzo/SP-tools}}.
Generally speaking, these features provide insight into amplitude (sound pressure level), envelope shape (temporal centroid), ``brightness" (spectral centroid), and ``noisiness" (spectral flatness).
See Caetano et al. \cite{caetano_audio_2019} for more in-depth information on timbre-related audio descriptors.

All audio features, except for temporal centroid which uses a 125ms window size, are computed using frame-based processing with a window size of 46.4ms with 75\% overlap.
Following recent suggestions \cite{schwar_multi-scale_2023}, spectral features are windowed using a flat-top window prior to the FFT and the resulting magnitude spectrum is compressed using the following function: $p(X) = \log(1 + X)$, 
where $X$ is a magnitude spectrum.
These modifications were shown to produce a smoother gradient for sinusoidal frequency estimation.
Audio feature time-series are summarized using the mean.

Building on Danielsen et al.~\cite{danielsen_effects_2015}, all features except for temporal centroid are extracted from two temporal segments within the same audio sample.
A short segment containing $N_t$ windows are selected from the onset to capture transient phase information and a longer segment containing $N_s$ windows are selected after the $N_t$ windows to capture the sustain/decay phase information.
The result is a seven dimensional audio feature space consisting of both timbral and dynamic features.

\subsubsection{Psychophysical Scaling}

Just as the equal-tempered scale enables transpositions of melodies between different keys, we seek a scaling that allows us to transpose sequences within the timbre/dynamic space defined by our audio feature extraction algorithm.
Dynamic features, computed as the root mean square (RMS), are converted to loudness, k-weighted, relative to full-scale (LKFS) \cite{international_telecommunication_union_algorithms_2006} by applying the following scaling function:

\begin{equation}
	s_{\text{LKFS}}(x_{\text{RMS}}) = -0.691 + 10\log_{10}(h(x_{RMS}))
\end{equation}

\noindent where $h(\cdot)$ is a K-weighting pre-emphasis filter.

Kazazis et al. \cite{kazazis_interval_2022} derived the following scaling function for spectral centroid:

\begin{equation}
    s_{\text{SC}}(x_{\text{SC}}) = -34.61x_{\text{SC}}^{-0.1621} + 21.2985
\end{equation}

\noindent where $x_{\text{SC}}$ is spectral centroid measured in hertz.

Temporal centroid is related to duration within our synthesis framework.
Schlauch et al. \cite{schlauch_duration_2001} found that perception of duration in damped sounds is dependent on frequency and timbre; however, derived power functions were all close to $d^{0.5}$ where $d$ is duration.
Accounting for the non-linear relationship between temporal centroid and duration of the exponential decay envelopes in our synthesizer, which was empirically determined by sampling envelopes, we derived the following psychophysical scaling for temporal centroid:

\begin{align}
    s_{\text{TC}}(x) = 0.03{x_{\text{TC}}^{1.864}}
\end{align}

To our knowledge, there is no literature investigating the perceptual scaling of spectral flatness; however, taking guidance from the Librosa documentation \cite{mcfee_librosa_2015}, spectral flatness is converted to a decibel scale: $s_{\text{SF}}(x) = 20\log_{10}(x_{\text{SF}})$.

Now equipped with a differentiable synthesizer $g(\cdot)$ and an audio feature extractor $f(\cdot)$ with perceptually informed scalings, we are ready to define a machine learning task for snare drum timbre remapping.

\subsection{Real-Time Timbre Remapping}

\begin{figure*}[htpb]
	\centering
		\includegraphics[width=1\textwidth]{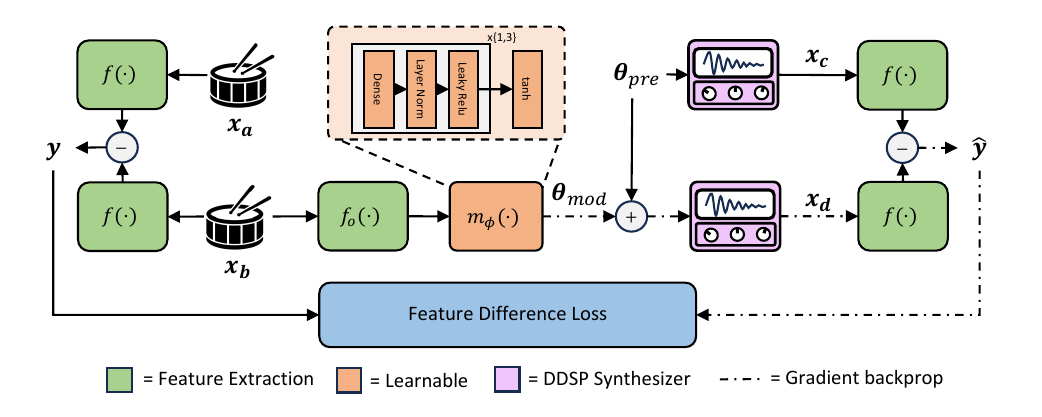}
	\caption{Real-time timbre remapping experiment overview. We learn a mapping network $m_{\phi}(\cdot)$ to predict synthesizer parameter modulations $\mathbf{\theta}_{mod}$ to create timbre analogies. Feature differences $\mathbf{y}$ are measured between two input sounds $(\mathbf{x}_a, \mathbf{x}_b)$ and $m_{\phi}(\cdot)$ learns to modulate a synthesizer preset $\mathbf{\theta}_{pre}$ to create a synthesized sound pair $(\mathbf{x}_c, \mathbf{x}_d)$ with a feature difference $\mathbf{\hat{y}}$. The feature difference loss measures the error between $\mathbf{y}$ and $\mathbf{\hat{y}}$. Real-time remapping is enabled by onset features $f_0(\cdot)$, which are measured on a short window of audio at a detected onset and are used as input to the mapping network.}
	\label{fig:overview}
\end{figure*}

We now consider how these concepts can be applied to the musical task of real-time control of a synthesizer.
The proposed system is inspired by the SP-Tools library, developed by percussionist and researcher Rodrigo Constanzo using FluComa \cite{tremblay_fluid_2022}, and recent work on percussive DMI control \cite{martelloni_real-time_2023}.
These works support real-time machine learning tasks using audio features extracted at detected onsets.
We explore here learning mappings between onset features and synthesizer parameter modulations for real-time timbral control.
A diagram overviewing this approach is provided in Figure \ref{fig:overview}.

We frame timbre remapping as a regression problem and use a data-driven approach to model the relationship between onset features and parameter modulations.
The goal is to estimate synthesizer parameter modulations $\mathbf{\hat{\theta}}_{mod}$ from short-term audio features $f_{o}(\mathbf{x})$, where $f_{o}$ is an onset feature extractor and  $\mathbf{x}$ audio from a single acoustic snare drum strike.
To do so we introduce a mapping function $m_{\phi}(\cdot)$ that outputs parameter modulations given onset features:

\begin{equation}
	\mathbf{\hat{\theta}}_{mod} = m_{\phi}(f_o(\mathbf{x}))
\end{equation}

\noindent where $\phi$ are the learnable model parameters.

Now our objective is to learn $\phi$ to estimate $\mathbf{\hat{\theta}}_{mod}$ to minimize the feature difference loss instead of directly optimizing $\mathbf{\hat{\theta}}_{mod}$.
To return to the concept of timbre analogies as a method for mapping, we construct a dataset of timbral sequences from an audio dataset of snare drum hits (e.g., snare drum hits extracted from a performance).
This is done by selecting a reference sample $\mathbf{x}_a$ from the dataset and then measuring the difference between that and every other sample in the dataset.
The reference sample acts as an anchor point in the dataset from which timbral/dynamic variations extend from and defines the sound for which the synthesizer preset is unmodulated (i.e., $\theta_{mod} = 0$).

Onset features are derived from a subset of features used in SP-Tools and are RMS, spectral centroid, and spectral flatness computed on a buffer of 256 samples after a detected onset.
Spectral features are computed on a magnitude spectrum and time domain samples are windowed with a Hann window prior to an FFT.
Onset detection is computed using an amplitude-based method derived on an implementation in the FluComa library \cite{tremblay_fluid_2022} (AmpFeature\footnote{\url{https://learn.flucoma.org/reference/ampfeature/}}) and used in SP-Tools.
All onset features are normalized to a $[0,1]$ range.

\section{Experiments}
Initial experimentation preceded the development of the concepts outlined in this paper and consisted of a manually-tuned mapping strategy where linear relationships between onset features and synthesis parameters were specified by a user on a user interface -- no machine learning involved.
This straightforward approach enabled everyday objects to be struck, transforming them into different elements of a synthetic drum kit.
Furthermore, we presented this version of the project at the \anonymize{Agential Insrtuments Workshop at the 2023 AI and Music Creativity} conference, where it was integrated into a project completed by two workshop participants.

This early success encouraged us to further develop this idea and address some of the main limitations: 1) manual-mapping is effective when there are relatively few features and synthesis parameters, but becomes unwieldy as complexity increases; 2) developing non-linear relationships with interdependencies between features is infeasible with manual-mapping; 3) creating mappings that lead to realistic graded timbral changes can be challenging.
The ideas presented in this paper represent our attempt to address some of the challenges.
Whilst we don't directly compare the manual mapping approach with the data-driven mappings in this paper, in the following subsections we present a series of numerical and musical experiments with the goal of providing insight into the efficacy and musical affordances of our timbre remapping design.

\subsection{Numerical Experiments}
Numerical experiments included training a set of different models to conduct real-time timbre remapping using an dataset of snare drum performances. 
We use subsets of the Snare Drum Dataset (SDSS) \cite{cheshire_snare_2020} to train and evaluate the real-time mapping model.
In total there are 48 unique performances and each recording contains between 50 to 120 (median 85) different hits.
We select audio from single microphone (an AKG-414 positioned on the top of the drum).
A full dataset for training a single model is one performance.

To generate timbral/dynamic analogies for each dataset, onset features and full audio features were computed for each individual drum and a reference drum sound $\mathbf{x}_a$ was selected using the median value of transient LKFS.
We found this feature correlated with strike velocity in a preliminary study and provides a reasonable centre point in the audio feature space to generate timbral/dynamic analogies from.
Testing and validation splits were created from each dataset using approximately ten percent of the samples, selected to roughly cover the dynamic range of the dataset.
Five different synthesizer presets were manually programmed to serve as the starting point for timbre analogies.
Combining the 48 performances with five different presets meant that 240 individual models were trained for each variation during experimentation.

\begin{table*}[htpb]
\centering
\caption{Feature Difference Errors}
\begin{tabular}{lcccccccc} 
\toprule
Feature & Preset & Direct & \multicolumn{2}{c}{Linear} & \multicolumn{2}{c}{MLP} & \multicolumn{2}{c}{MLP LRG} \\
\cmidrule(lr){4-5}
\cmidrule(lr){6-7}
\cmidrule(lr){8-9}
\multicolumn{3}{c}{} & 256 & 2048 & 256 & 2048 & 256 & 2048\\
\midrule
$LKFS_T$ & $19.6 \pm 4.7$ & $0.473 \pm 1.1$ & $1.196 \pm 0.8$ & $\mathbf{0.479 \pm 0.6}$ & $0.962 \pm 1.0$ & $0.743 \pm 1.0$ & $1.016 \pm 1.0$ & $0.808 \pm 1.1$ \\
$LKFS_S$ & $60.8 \pm 21$ & $1.244 \pm 1.3$ & $2.662 \pm 1.7$ & $2.751 \pm 1.8$ & $2.131 \pm 1.6$ & $2.104 \pm 1.6$ & $2.083 \pm 1.6$ & $\mathbf{2.081 \pm 1.5}$ \\
$SC_T$ & $12.7 \pm 1.4$ & $0.120 \pm 0.1$ & $0.166 \pm 0.1$ & $0.163 \pm 0.1$ & $0.129 \pm 0.1$ & $\mathbf{0.125 \pm 0.1}$ & $0.130 \pm 0.1$ & $0.133 \pm 0.1$ \\
$SC_S$ & $12.8 \pm 1.4$ & $0.221 \pm 0.1$ & $0.228 \pm 0.1$ & $0.231 \pm 0.1$ & $\mathbf{0.223 \pm 0.1}$ & $0.225 \pm 0.1$ & $0.230 \pm 0.1$ & $0.233 \pm 0.1$ \\
$SF_T$ & $48.4 \pm 45$ & $1.392 \pm 2.0$ & $3.211 \pm 2.1$ & $2.091 \pm 2.1$ & $2.307 \pm 1.8$ & $\mathbf{1.747 \pm 1.8}$ & $2.354 \pm 1.9$ & $1.953 \pm 2.0$ \\
$SF_S$ & $34.2 \pm 46$ & $2.075 \pm 2.7$ & $4.284 \pm 2.8$ & $4.124 \pm 2.5$ & $\mathbf{3.645 \pm 2.6}$ & $3.662 \pm 2.6$ & $3.799 \pm 2.5$ & $3.881 \pm 2.6$ \\
$TC$ & $22.7 \pm 22$ & $1.907 \pm 5.0$ & $2.771 \pm 5.0$ & $2.593 \pm 4.9$ & $2.264 \pm 4.9$ & $2.222 \pm 4.9$ & $2.215 \pm 4.9$ & $\mathbf{2.164 \pm 4.9}$ \\
\bottomrule
\end{tabular}
\begin{tablenotes}
\small
\item LKFS: Loudness, K-Weighted, relative to full-scale; SC: Spectral Centroid; SF: Spectral Flatness; TC: Temporal Centroid
\item T: Transient phase; S: Sustain phase.
\item Lower values are better for all values and results in bold highlight the best modeling approach for each feature.
\end{tablenotes}
\label{tab:results}
\end{table*}

\vspace{1mm}
\subsubsection{Models and Training}
Three different model variations were included for experimentation. 
Two models based on the multi-layer perceptron (MLP) used by Engel et al. \cite{engel_ddsp_2020} were included. 
One with a single layer containing 32 hidden units (590 parameters) and a larger model with three layers of 64 hidden units (9.5k parameters).
A linear model was also included for experimentation, which reflects the mapping capability of the aforementioned manual-mapping method.

Two window sizes for onset features were also included, one with short-term features of 256 samples, and one with a larger window of 2048 samples.
The larger window ($\approx$ 43ms at 48kHz) would have too much latency for real-time percussion performance, which has an upper perceptual threshold of about 10ms \cite{jack_action-sound_2018}.
We include this larger window to investigate the benefit of providing more temporal context during training.
Parameters were also directly optimized to match differences (i.e., no modeling) to evaluate the effectiveness of the feature difference loss and provide an upper bound on performance.

Each model was trained for 250 epochs using an Adam optimizer and the learning rate was halved if validation loss did not improve for 20 epochs.
To prevent over utilization of oscillator frequency parameters, modulations for those two controls were damped by a factor of 1e-3.
Training a model takes under 2 minutes on a NVIDIA GeForce RTX 2080 Ti GPU and about 13 minutes on the CPU of a MacBook Pro M1.
A full listing of hyperparmaters and model details are listed on the supplementary website.

\subsubsection{Results}

Metrics reporting how accurately each model variant was able to match audio feature differences are shown in table \ref{tab:results}.
All results were computed using samples from testing datasets and results are summarized with mean and standard deviation across all SDSS performance and preset pairs.
The preset column shows the feature difference error computed against the preset before any optimization to provide a performance baseline -- this would represent a one-shot triggering scenario.
The direct optimization results performed the best across all features, which is as expected since no model is being trained to estimated parameters.
While the non-linear MLP models tend to provide better feature matching capabilities across most features, they are not significantly better than the best linear model.
These results show that we were able to relatively accurately learn to map synthesizer parameter changes using this approach compared to the baseline direct optimization and that these relationships can be modeled within a single snare drum performance with relatively simple models.

\subsection{Musical Experiments}
Deruty et al. \cite{deruty_development_2022} emphasize working alongside musicians in the development AI-tools and highlight the importance of creating usable prototypes that function within a musicians typical workflow.
To facilitate experimentation within the intended musical context, an audio plug-in was developed to perform real-time timbre remapping.
The only difference between the plug-in and training is that the audio feature extraction algorithms and synthesizer was re-written in C++ (as opposed to Python) and a rolling feature normalizer was added to ensure that input features were in the correct $[0,1]$ range.
The plug-in, source code, and recordings from the musical experiment are available on the supplementary website. 

We conducted an informal session with professional drummer Carson Gant to record musical examples to accompany this paper and help situate this work within the practice of a groove-based drummer.
The goal of this session was provide initial feedback of our approach within a musical context and is not intended to replace a formal user study, which the authors plan to conduct at a later date.
Carson provided short recordings of performances on two different snare drums with and without dampening, which were used to train models ahead of our session.
An important distinction between the recordings received from Carson and the SDSS dataset is that Carson played a much wider range of gestures including buzz roles and rim clicks.

After playing for a period of time, Carson remarked ``there is some subtleness to it where you're not getting one-shotted\footnote{Referring to the effect of re-triggering a recorded sample repeatedly, sometimes called a ``machine-gun effect" \cite{fagerstrom_one--many_2021}}, there are subtle changes to it ... it's nice to hear, it's reacting ... it's just figuring out how to play it and what causes it to trigger [or not]".
This statement points to both a success of the timbre remapping in creating subtle variations and a limitation of relying on onset detection.
Carson's statement reflects findings by Jack et al. \cite{jack_rich_2017}, who observed percussionists reducing their gestural language when confronted with the bottleneck of discrete onset detection in a percussive DMI.
It is worth noting that the setup in our session, which used a dynamic microphone as input, represents a challenging scenario for onset detection and could likely be significantly improved with the introduction of a drum trigger (p.c., Rodrigo Constanzo).

Carson played several different models, remarking that some ``felt more reactive" or ``were triggering better."
This was interesting as the onset detection and triggering is separate from the mapping model.
This points to a perceptual connection between the variation in sound produced and a sensation of reactiveness.
A feeling of less reactivity was particularly salient in presets that contained higher levels of filtered noise.
Small adjustments in high-pass filter parameters (cut-off and q) created perceptually significant changes and minor variations in the input features tended to feel over-emphasized.
Carson mentioned that this resulted in a sensation of randomness, although also commented that playing on the edge of the drum resulted in one sound and in the middle another, suggesting that macro control worked well, but granular control over noise was marginal.
This variability could also be attributed in part to the feature normalization in the audio plug-in, which would update over time and could cause outputs to change over time.

\section{Discussion}

\subsection{Differentiable Timbre Space}

The timbre space has proved an enduring concept for control of sound synthesizers.
The marriage of DDSP with timbre space in this work offers a novel perspective, enabling the direct learning of synthesis parameters from audio examples, circumventing the need for supervision on parameters \cite{vaillant_interpolation_2023} or generative training on large datasets \cite{nistal_drumgan_2020}.
Expressing perceptual knowledge directly within our DDSP training algorithm allowed us to explicitly specify the musical concepts that we deemed important for the task at hand.
In this case, we highlighted the importance of timbral differences between events in a musical phrase by using a feature difference loss.
This enabled the efficient training of lightweight models capable of performing real-time timbre remapping.
However, representing a sound as a point in a multidimensional and numerical timbre space also involves a reduction -- a timbral bottleneck -- similar to the gestural bottleneck introduced through onset detection \cite{jack_rich_2017}.
Bottlenecks in our design had implications that were reflected in our musical experiment and reveal avenues for future investigation.

\subsection{Limitations}

The proposed feature difference loss was designed under the assumption that timbre variations can be represented as vectors in a computational timbre space and that vectors with the same direction (but different origins) will be perceived similarly. 
Previous research on timbre analogies and scaling of timbre-related audio descriptors supports this assumption, and our musical experiment offers an initial practice-based evaluation of its perceptual relevance. 
Further investigation into the perceptual relevance of the proposed method in psychoacoustical and musical contexts will be both enlightening and important for future development in this direction.
Additionally, the choice of reference in the feature difference loss has implications on the end result and future work can explore different formulations such as dynamic references that are updated based on shorter musical phrases.

While DDSP offers numerous benefits, it also introduces some unique challenges.
The difficulty of optimizing frequency with respect to audio loss functions is well-known \cite{turian_im_2020, masuda_improving_2023}.
Our training scenario avoided the need to directly learn frequency; however, we observed uninformative gradients with respect to frequency parameters.
The impact of uninformative gradients meant model weight initialization had a large impact on training and necessitated the use of frequency parameter dampening to mitigate bad solutions.
Despite these challenges, and in light of the clear aforementioned benefits and recent insights in DDSP optimization \cite{hayes_sinusoidal_2023, schwar_multi-scale_2023}, we feel that continued research in this direction is merited.
Future work comparing non-differentiable approaches \cite{vaillant_interpolation_2023} would also be worthwhile.

\subsection{Opportunities}

Beyond the real-time percussive timbre remapping application presented in this work, there are numerous applications of timbre remapping with DDSP.
We highlight a few here.
A simple extension of our current work is explore the benefits of exposing direct parametric control over timbral features.
In our case study, onset features are used as input to the parameter mapping neural network.
However, there is no reason why values from an external controller couldn't be mapped to these.
One application that we have already started to explore is mapping MIDI and MPE values from a controller like the Ableton Push\footnote{\url{https://www.ableton.com/en/push/}}.
This could enable more nuanced timbral control over synthesis parameters in finger-drumming and other controller-based performance contexts.
Furthmore, parametric timbre control could be used in sound design applications or to create stimuli for perceptual studies where independent control over individual features is beneficial and typically relies on additive synthesis \cite{kazazis_interval_2022}.
Creation of meaningful variations in sounds is an another active area of research for drum one-shots \cite{fagerstrom_one--many_2021} and sound effects for video games \cite{siddiq_real-time_2015}.
Timbre variations could be learned in a data-driven manner from sample libraries, for instance, by creating differentiable implementations of procedural synthesis methods \cite{menexopoulos_state_2023}.

\section{Conclusion}
In this paper we have explored how timbre analogies and differentiable audio synthesis can be leveraged together for the task of timbre remapping.
Specifically, we sought to map subtle timbral changes from acoustic instruments onto controls for a synthesizer, motivated by musical contexts where timbre is a primary vehicle for expression.
By expressing synthesis and feature extraction algorithms differentiably, and through the use of our proposed feature difference loss function, we showed how we could learn to adjust synthesis parameters to match timbral and loudness feature sequences.
Importantly, we matched differences, as opposed to absolute values audio features, which emphasized the importance of trajectories in timbre space and enabled remapping.
This was shown in a concrete example that explored real-time remapping from acoustic snare drum performances to a differentiable drum synthesizer inspired by the Roland TR-808.

\section{Acknowledgments}
\anonymize{
This work is supported by the UKRI through the Centre for Doctoral
Training in Artificial Intelligence and Music (EP/S022694/1) and a
UKRI Frontier Research grant (EP/X023478/1).
We would like to thank Carson Gant for his help and feedback during the performance session.
We would also like to thank Rodrigo Constanzo for his inspiring work on the SP-Tools library and for the insightful conversations. Thank you to all the NIME reviewers for their valuable feedback.
}

\section{Ethical Standards}
All research carried out as a part of this work was conducted solely by the authors.
Material related to the musical session with Carson Gant, including quotes and video recordings on the supplemental website, are included with his permission.
We acknowledge the potential for machine learning technology such as this to misused, for instance in the creation of fake or misleading content.
We've endeavored to use small data and models to mitigate this risk and to enable musicians to utilize this technology themselves.

\bibliographystyle{abbrv}
\bibliography{final-ref} 

\end{document}